\documentclass[osajnl,twocolumn,showpacs,superscriptaddress,10pt]{revtex4-1} 
\usepackage{amsmath,amssymb,graphicx}
\begin{document}

\title{Generation of correlated photon pairs in micro/nano-fibers}

\author{Liang Cui}
\author{Xiaoying Li}\email{Corresponding author: xiaoyingli@tju.edu.cn}
\author{Cheng Guo}
\affiliation{ College of Precision Instrument and Opto-electronics Engineering, Tianjin University, Key Laboratory of Optoelectronics Information Technology of Ministry of Education, Tianjin 300072, China}

\author{Y. H. Li}
\affiliation{Joint Institute for Measurement Science, Tsinghua University, Beijing 100084, China}
\affiliation{State Key Laboratory of Precision Measurement Technology and Instruments, Tsinghua University, Beijing 100084, China}
\author{Z. Y. Xu}
\affiliation{Joint Institute for Measurement Science, Tsinghua University, Beijing 100084, China}
\author{L. J. Wang}
\affiliation{Joint Institute for Measurement Science, Tsinghua University, Beijing 100084, China}
\affiliation{State Key Laboratory of Precision Measurement Technology and Instruments, Tsinghua University, Beijing 100084, China}

\author{Wei Fang}
\affiliation{Department of Optical Engineering, Zhejiang University, Hangzhou 310027, China}

\begin{abstract}
We study the generation of correlated photon pairs via spontaneous four wave mixing in a 15 cm long micro/nano-fiber (MNF). The MNF is properly fabricated to satisfy the phase matching condition for generating the signal and idler photon pairs at the wavelengths of about 1310 and 851 nm, respectively. Photon counting measurements yield a coincidence-to-accidental ratio of 530 for a photon production rate of about 0.002 (0.0005) per pulse in the signal (idler) band. We also analyze the spectral information of the signal photons originated from the spontaneous four wave mixing and Raman scattering. In addition to discovering some unique feature of Raman scattering, we find the bandwidth of the individual signal photons is much greater than the calculated value for the MNF with homogeneous structure. Our investigations indicate the MNF is a promising candidate for developing the sources of nonclassical light and the spectral property of photon pairs can be used to non-invasively test the diameter and homogeneity of the MNF.

\end{abstract}


\maketitle 

The quantum light sources, such as squeezed states and correlated photon pairs, are the key resource of quantum information technology~\cite{Gisin02RMP,Braun05}.
The most employed ways for generating the quantum light are the optical parametric processes in $\chi^{(2)}$-based nonlinear crystals \cite{Shih03RPP,Reid09} and in $\chi^{(3)}$-based optical fibers or atoms~\cite{LJWang01JOB,Fiorentino02PTL,Fan05OLa,Guo12,Boyer08}.
To make the quantum technologies really practical, many efforts have been made on micro-minimizing and integrating the sources of quantum light \cite{Sharping06oe,Lin06OL}.

The micro/nano-fibers (MNFs) have been extensively investigated owing to their potential usefulness as building blocks in future micro- or nanometer-scale photonic components or devices~\cite{Tong10BOOK}. In addition to the characteristics of small footprints, relatively large evanescent field, and strong near-field interaction, the MNFs also have the properties of the miniaturized modal area, engineerable dispersion, large available length, low wave-guiding loss, and easy connection to a standard fiber optical system. More applications of MNF, including optical nonlinear effects and atom manipulation, have been explored in recent years~\cite{Birks00OL,Foster04OE,Wang12,Balykin04}. However, the application of MNF in generating quantum light has not been reported yet.

In this letter, we demonstrate a source of quantum correlated photon pairs via spontaneous four wave mixing (SFWM) in a 15 cm long MNF for the first time.
The MNF is drawn from SMF-28 single mode fiber (SMF).
The wavelengths of the photon pairs are about 1310 and 851 nm, respectively, and the measured coincidence-to-accidental ratio is about 530 when the production rate of signal (idler)
photon is about 0.002 (0.0005) per pulse.
Comparing with the photonic crystal fiber (PCF) based source of photon pairs with large frequency detuning~\cite{Rarity2005OE},
it is easier to integrate the MNF based source into the existing fiber system with lower insertion loss.

For the SFWM, when a pulsed pump is launched into the $\chi^{(3)}$-based optical fiber, two pump photons at wavelength $\lambda_p$ scatter into a pair of signal and idler photons at wavelengths $\lambda_s$ and $\lambda_i$, respectively.
The SFWM process is governed by energy conservation and the phase matching condition:
\begin{eqnarray}\label{eq-ec}
2/\lambda_p=1/\lambda_s+1/\lambda_i,
\end{eqnarray}
\begin{eqnarray}\label{eq-pmc}
2k_p-k_s-k_i-2\gamma P_p=0.
\end{eqnarray}
Here, $k_p$, $k_s$ and $k_i$ are the propagation constants of the pump, signal and idler field, respectively; $2\gamma P_p$, with $\gamma$ and $P_p$ respectively denoting the nonlinear coefficient of the fiber and the peak power of the pump, is the self phase modulation term.

\begin{figure}[tbp]
  \centering
  \includegraphics[width=8.1 cm]{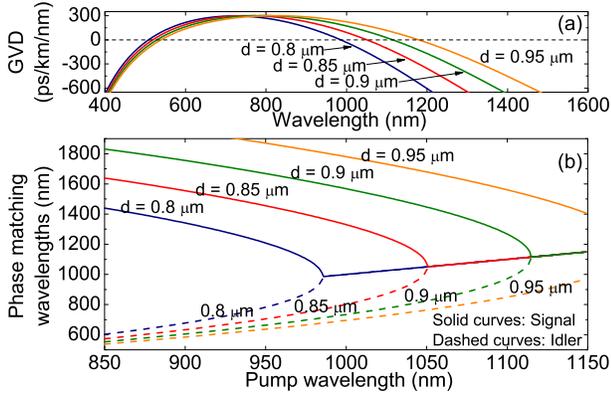}
  \caption{(Color online) Calculated results of (a) group velocity dispersion (GVD) and (b) SFWM phase matching curves for the MNFs with different diameters.}
  \label{FIGPM}
\end{figure}

In general, the dispersion property of an air-cladding MNF is determined by its diameter $d$. To properly fabricate the MNF, we first compute the propagation constants and the group velocity dispersion (GVD) of the fundamental mode for the MNFs with diameters $d$ of 0.8, 0.85, 0.9 and 0.95 $\mu m$, respectively, by using the standard procedure for analyzing the step-index optical fibers \cite{Tong10BOOK}.
As shown in Fig. \ref{FIGPM}(a), the GVD varies with $d$ and each kind of MNF has two zero-GVD wavelengths within the plotted range.
According to the computed propagation constants and Eqs. (\ref{eq-ec}) and (\ref{eq-pmc}), we then calculate the phase matching wavelengths of SFWM for the pump wavelength varying from 850 to 1150 nm, as shown in Fig. \ref{FIGPM}(b).
We note that in the calculation, the self phase modulation term in Eq. (\ref{eq-pmc}) is omitted due to its small magnitude.

Fig. \ref{FIGPM}(b) shows that the frequency detuning between the pump and signal (idler) photons is very small when the pump wavelength is close to the longer zero-GVD wavelength and in the normal dispersion region. However, the detuning dramatically increases with the wavelength difference between the pump and the zero-GVD when the pump wavelength is in the anomalous dispersion region. Clearly, for a given pump source, one can generate photon pairs with desired wavelengths by properly fabricating the MNF to engineer its dispersion property.
For example, using a laser centering at $\sim$1040 nm, photon pairs with one photon in the 1310 nm O-band and the other in the 850 nm band can be realized if the diameter of MNF is about 0.9 $\mu m$.
The reason we are interested in this kind of photon pairs is twofold: i) the 1310 nm photon is compatible with the existing fiber systems, while the 850 nm photon has high detection efficiency by using the commercially available single photon detectors (SPD), so the photon pairs are suitable for building heralded single photon sources in telecom bands with high efficiency; ii) the noise photon originating from Raman scattering (RS) can be significantly suppressed at room temperature due to the large detuning between the pump and signal (idler) photons.

\begin{figure}[tbp]
  \centering
  \includegraphics[width=8.2 cm]{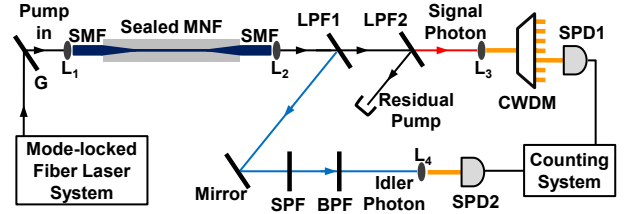}
  \caption{(Color online) Experimental setup. See text for details.}
  \label{FIGSETUP}
\end{figure}

The experimental setup for generating photon pairs in MNF is depicted in Fig. \ref{FIGSETUP}.
A homemade mode-locked laser system consisting of a seed oscillator and a one-stage amplifier is employed as the pump source. The system is based on the Yb-doped fibers, so it is compact in size.
The repetition rate and pulse duration of the output pulse train are $\sim$62.56 MHz and $\sim$250 fs, respectively.
The output spectrum with a bandwidth of about 20 nm is centering at 1041 nm, and the average output power can reach 1.5 W.
By adjusting the transmission grating G, we are able to tailor the pump with a specified central wavelength.

To obtain photon pairs at about 1310 and 850 nm, respectively, we fabricate a 15 cm long MNF with diameter of about 0.9 $\mu$m using a taper-drawing workstation.
During the drawing process, the SMF with a core diameter of 9 $\mu$m is heated by a millimeter-size flame of a hydrogen-oxygen mixture that is regulated by separating gas-flow controllers to guarantee a constant gas flow and uniform heating. In the meantime, the SMF is pulled by two computer-controlled translation stages. The heating region increases gradually and precisely with taper extension until the designed length and diameter is obtained. To prevent dust contamination and to ease the handling of MNF, we seal the MNF and the adiabatically tapered region
into a plastic housing.
We also measure the total transmission efficiency of the MNF, and
the results are: 75\% at 1310 nm, 63\% at 850 nm, and 48\% at 1040 nm. The decreased efficiencies at 850 and 1040 nm may be caused by the loss of the higher-order propagation modes in the adiabatic taper.

Before coupling the pump into the MNF from the cleaved SMF end, the aspheric lens L$_1$ is used to focus the light passing through the grating G.
The bandwidth and central wavelength of the pump can be tuned by moving the positions of L$_1$ and the SMF end of the MNF. In the experiment, the central wavelength of pump launched into the MNF can be changed within the range of 1031-1051 nm, and the full width at half maximum (FWHM) of pump is about 1.5 nm.

Since the efficiency for generating photon pairs in MNF is very low, only about 0.03 pairs per pulse are generated by 10$^8$ pump photons per pulse, an isolation to pump of in excess of 110 dB is required for reliably detecting the signal and idler photons. Hence, after passing the output of MNF through an aspheric lens L$_2$, we separate the signal, idler and pump photons by using two interferometric long pass filters (LPFs).
The idler photons are reflected by LPF1 (cut-off wavelength: 950 nm) and the residual pump is reflected by LPF2 (cut-off wavelength: 1200 nm). Only the signal photons can pass through LPF2.
To further enhance the isolation of the pump photons and restrict the spectra of the detected photon pairs, we insert a short pass filter (SPF, cut-off wavelength: 950 nm) and a band pass filter (BPF) in the idler band, while let the output of LPF2 pass through a coarse wavelength division multiplexer (CWDM). The central wavelength and bandwidth of the BPF can be selected depending on experimental requirements. The CWDM has 18 channels, whose central wavelengths range from 1270 to 1610 nm and the spacing between adjacent channels is 20 nm. For each channel, the transmission spectrum is super-Gaussian shaped and the corresponding 0.5dB bandwidth is about 18 nm.

The signal and idler photons are detected by SPD1 and SPD2, respectively.
SPD1 (id200, IDQ) working in the gated Geiger mode is based on the InGaAs/InP avalanche photon diode, while SPD2 (SPCM-AQRH, Perkin-Elmer) working in the active quenching mode is based on the Si avalanche photon diode.
For the counting measurements of individual signal photons, SPD1 is triggered by a 6.95 MHz TTL pulse train, 1/9 division of the repetition rate of the laser system; while for the two-fold coincidence counting measurements of the signal and idler photons, the SPD1 is triggered by the detection events of SPD2. The quantum efficiencies of SPD1 and SPD2 are about 10\% and 40\%, respectively.

\begin{figure}[tbp]
  \centering
  \includegraphics[width=8.5 cm]{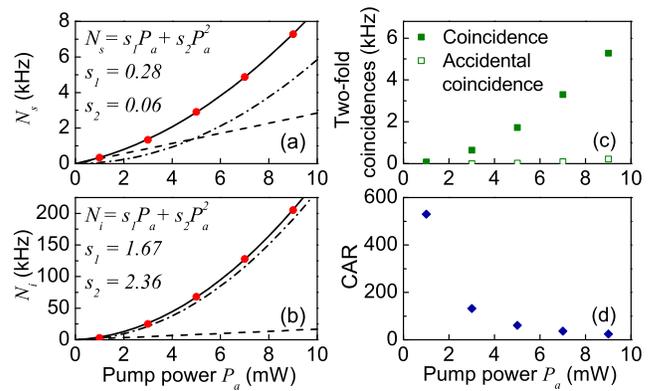}
  \caption{(Color online) Counting rate of (a) individual signal and (b) idler photons, $N_s$ and $N_i$, versus the pump power $P_a$ (solid circles). The solid curve is the fitting of second-order polynomial function $N_{s(i)}=s_1P_a+s_2P_a^2$, with the dashed line and dot dashed curve respectively representing the linear and quadratic parts. (c) Two-fold coincidence counting rates and (d) CAR versus $P_a$. The error bars of the data are within the size of the data points.}
  \label{FIGCOIN}
\end{figure}

We first observe the generation of the individual signal and idler photons via SFWM. In this experiment, signal photons are transmitted through the channel of CWDM having the central wavelength of 1310 nm. By adjusting the central wavelength of pump $\lambda_p$ and recording the counting rate of SPD1, we find that for the pump with a fixed power, the counting rate of SPD1 $N_s$ changes with the variation of $\lambda_p$ and are quite high for the case of $\lambda_p=1031.8$ nm. Fig. \ref{FIGCOIN} (a) plots $N_s$ as a function of the average power of pump $P_a$ for $\lambda_p=1031.8$ nm.
Fitting the measured data with the second-order polynomial function $N_{s}=s_1P_a+s_2P_a^2$, where the fitting coefficients $s_1$ and $s_2$ are respectively determined by the strengths of RS and SFWM \cite{Li05c}, we find the quadratic part $s_2P_a^2$ dominates when $P_a$ is greater than 5 mW.
The results indicate that the phase matching condition of SFWM is satisfied.

According to energy conservation, we then set the central wavelength of the Gaussian shaped BPF in the idler band at 851 nm, whose FWHM is about 9 nm.
Fig. \ref{FIGCOIN} (b) shows the counting rate of SPD2 $N_i$ obtained by varying the pump power. By fitting the measure data with the function $N_{i}=s_1P_a+s_2P_a^2$, one sees that the contribution of linear part $s_1P_a$ is very small and almost all the counting rates in the idler band originate from SFWM.

To verify the quantum correlation of photon pairs, we measure the coincidence and accidental coincidence counting rates by recording the two-fold coincidences of SPD1 and SPD2 for the signal and idler photons originated from the same and adjacent pulses, respectively, and calculate the  coincidence-to-accidental ratio (CAR) as well. As shown in Figs. \ref{FIGCOIN} (c) and \ref{FIGCOIN} (d), for the case of $P_a$ = 1 mW (9 mW), the measured coincidence counting rate and CAR are 74 Hz (5.3 kHz) and 530 (25), respectively, showing the strong correlation between the signal and idler photons.
After taking the total detection efficiency of signal and idler photons into account, which are about 2\% and 10\%, respectively, it is straightforward to deduce that for the CAR of 530 (25), the corresponding total production rates in the signal and idler bands are about 0.002 (0.06) and 0.0005 (0.03)  photons per pulse~\cite{Rarity2005OE}, respectively.

One may notice that for the given pump with a certain power, the production rates of individual signal and idler photons are not equal even if the difference of the FWHM of the detected signal and idler photons is considered.
This is because of the background noise photons originated from RS.
The value of the Bose population factor of optical phonon in the idler band is much smaller than that in the signal band~\cite{Li05c}, so the contribution of RS in the idler band is much less than that in the signal band.

\begin{figure}[tbp]
  \centering
  \includegraphics[width=8.5 cm]{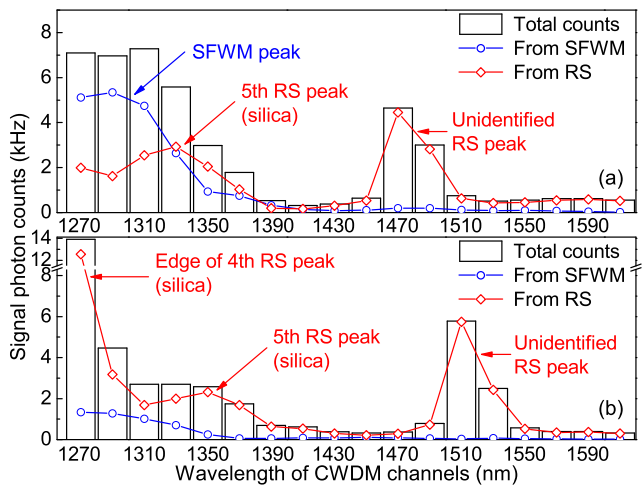}
  \caption{(Color online) Counting rate of signal photons in each CWDM channel, including the total rate (columns) and the rates originate from the SFWM (hollow circles) and the RS (hollow diamonds). (a) and (b) are obtained for $\lambda_p=1031.8$ and $\lambda_p=1050.3$ nm, respectively.}
  \label{FIGSIGNAL}
\end{figure}

To investigate how the RS in the signal band affects the purity of photon pairs, for the photons emerging from each channel of the CWDM, we measure the counting rate of SPD1 by varying the pump power and fitting the measured data with the function $N_{s}=s_1P_a+s_2P_a^2$. By doing so, in each channel, we are able to obtain the total counting rate and to identify the proportions of photons originated from RS and SFWM, respectively. Fig. \ref{FIGSIGNAL} (a) shows the results for $P_a=9$ mW.
We note that the results in Fig. \ref{FIGSIGNAL} (a) have been normalized by using the transmission efficiency of the 1310 nm channel,
because the efficiencies of different channels of the CWDM are not equal.
One sees that the peak of the SFWM photons sits around 1310 nm and has a FWHM greater than 80 nm.
Since the calculated results indicate that the FWHM of the signal photons generated from a 15 cm long homogeneous MNF via SFWM should be about 7 nm,
the broadened SFWM spectrum in Fig. \ref{FIGSIGNAL} (a) implies that the diameter of the MNF is inhomogeneous \cite{Cui12PRA}.
Moreover, we find that for the photons contributed by RS, there are two peaks: one is at about 1330 nm, and the other is at about 1470 nm ($\sim$87 THz away from the pump). The former is identified as the 5th-order RS peak of silica~\cite{Rarity2005OE}. However, we can not identify the latter, because this is different from the spectrum of RS observed in conventional optical fibers and photonic crystal fibers~\cite{Agrawal,Rarity2005OE}.

As a further characterization of the RS peak at $\sim$87 THz away from the pump, we repeat the measurement for the signal photons in each channel of CWDM when $\lambda_{p}$ is tuned to 1050.3 nm, which is about 3.4 THz away from 1031.8 nm.
The results are shown in Fig. \ref{FIGSIGNAL} (b), which are also obtained for $P_a=9$ mW. Comparing with Fig. \ref{FIGSIGNAL} (a), one sees that the peaks of RS move to the longer wavelength side, and the edge of the 4th-order RS having a extremely large counting rates just comes into the wavelength range of CWDM. In contrast, the peak of the photons originated from SFWM moves to the shorter wavelength side, which agrees with the theoretical expectation (see Fig. \ref{FIGPM} (b)).
Moreover, from Figs. \ref{FIGSIGNAL}(a) and (b), one sees that the translation of the peaks originated from the linear part of the fitting polynomial $N_{s}=s_1P_a+s_2P_a^2$ is parallel to the frequency shift of pump.
The results further confirm the unidentified peak of the scattered photons is originated from RS. Further investigation of RS in MNFs, which we are presently conducting, is necessary.

In conclusion, we have demonstrated the generation of correlated photon pairs in the 15 cm long MNF via SFWM.
The CAR of 530 for a signal (idler) photon production rate of about 0.002 (0.0005) per pulse is obtained in the correlation measurements.
Moreover, spectral analysis of the signal photons in the wavelength range of 1270-1610 nm reveals that the bandwidths of the photon pairs are much greater than the theoretically expected value due to the inhomogeneity of the MNF,
and experimental observation illustrates that the spectrum of RS in MNF is different from that in conventional optical fibers and PCFs.
Our investigation shows that the MNF is a promising candidate for developing the sources of quantum light in micro- or nanometer-scales, and the spectral property of photon pairs can be used to non-invasively test the diameter and homogeneity of the MNF.

This work was supported in part by the National Natural Science Foundation of China (Nos. 11304222 and 11074186), and the State Key Development Program for Basic Research of China (No. 2010CB923101).


\end{document}